\documentclass[twocolumn,aps,superscriptaddress,showpacs]{revtex4}
\usepackage{amsfonts}
\usepackage{graphicx}
\usepackage{amssymb,amsmath}
\newcommand {\be}{\begin{equation}}
\newcommand {\ee}{\end{equation}}
\newcommand {\bef}{\begin{figure}}
\newcommand {\eef}{\end{figure}}
\newcommand{\bea}{\begin{eqnarray}}
\newcommand{\eea}{\end{eqnarray}}
\begin{document}
\frenchspacing
\title{Barab\'asi Queueing Model and Invasion Percolation on a tree}
\author{A. Gabrielli$^1$, G. Caldarelli}
\affiliation{SMC, INFM-CNR, Dipartimento di Fisica Universit\`a 
di Roma ``Sapienza", Piazzale A. Moro 2, 00185 Rome, Italy,\\
ISC, CNR, Via dei Taurini 19, 00185 Rome, Italy.} 
\begin{abstract}
In this paper we study the properties of the Barab\'asi model of
queueing under the hypothesis that the number of tasks is steadily
growing in time. We map this model exactly onto an Invasion Percolation
dynamics on a Cayley tree. This allows to recover the correct 
waiting time distribution
$P_W(\tau)\sim \tau^{-3/2}$ at the stationary state (as observed in 
different realistic data) and also to characterize
it as a sequence of causally and geometrically connected bursts of activity. 
We also find that the approach to stationarity is very slow.

\end{abstract}
\maketitle 

Queueing theory \cite{CX61,GH83,BB05} describes a wide range of human
dynamical behaviors \cite{reynolds03,anderson03}.  Most of traditional
models lead to exponential {\em waiting time distributions} (WTD)
$P_W(\tau)\sim \exp(-\tau/\tau_0)$ for the tasks in the queue.
Recently, motivated by observations related to web browsing, email
communications and ordinary mail correspondence \cite{Vazquez06}, much
attention has been payed to priority driven queueing models generating
power-law WTD $P_W(\tau)\sim\tau^{-\alpha}$ for the tasks.  
In this paper we study a particular version of one of the latter kind: 
the Barab\'asi queueing model (BQM) \cite{bara1,bara2}
In our version of BQM at each
time step the task with highest random priority is always executed and
replaced in the queue by a constant number $m\ge 2$ of new tasks with
random priorities.  This process can be mapped exactly onto an
Invasion Percolation (IP) dynamics \cite{WW83} on a Cayley tree
\cite{IP-Cayley} with a series of advantages.  
Firstly we can characterize the task list dynamics
through the WTD at the stationary state. Secondly we show 
that its general evolution is composed by a sequence of geometrically
and causally connected burst of activities (task avalanches) with
scale-invariant size distribution. Thirdly, we can 
study the dynamics out of stationarity and we show that the approach
to it is very slow. Finally, it permits to simply generalize the results 
in to the case of time-varying $m$.
In the general BQM \cite{bara1} 
one starts with an initial list (i.e. queue) of $n_0\ge 2$ tasks. At every
time-step $t$ one of these tasks is executed and replaced by $m(t)$
other new tasks.  For constant $m(t)=1$, the queue length remains
constant.  The execution rule at each time-step is given by fixing a
random priority index $x_i\in[0,1]$ for each task in the queue and
then executing with a probability $p\le 1$ the task with the highest
priority and with a probability $(1-p)$ a randomly chosen task.  The
related problem for general $0\le p\le 1$ and $m=1$ has been analyzed 
and solved in \cite{Vazquez05,GC07}.  In the purely extremal (i.e. when $p=1$)
case with a variable queue length, the behavior of $P_W(\tau)$ differs
strongly from the previous case.  In \cite{GL06} it has been studied
the case in which at each time-step there is a probability
$\mu\le 1$ to execute the highest priority task, while a new task is
added to the list with another probability $\nu\le 1$. For $\mu=\nu=1$
the above case of conserved queue length is recovered.  
If at least one among $\mu$ and $\nu$ is strictly smaller than
$1$, the list length instead varies in time.  Depending whether
$\mu>\nu$, $\mu=\nu<1$ or $\mu<\nu$ the WTD $P_W(\tau)$ at the
stationary state changes the asymptotic behavior.  In particular for
$\mu=\nu<1$ all tasks are executed with $P_W(\tau)\sim
\tau^{-3/2}$ with no upper cut-off, while for $\mu<\nu$, the mean queue
length grows linearly in time, and one can show that asymptotically
for $t\rightarrow\infty$ all tasks with priority index $x_i<
(1-\mu/\nu)$ are never executed staying forever in the queue. Instead
tasks with $x_i\ge (1-\mu/\nu)$ are executed with a WTD coinciding
with the one for $\mu=\nu<1$.
\begin{figure}[t!]
\includegraphics[width=8.5cm]{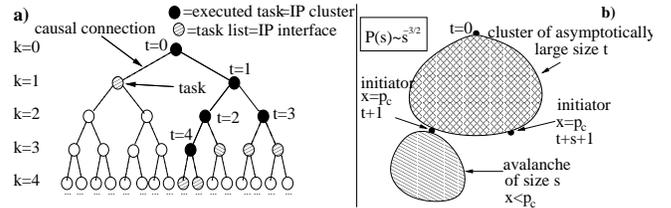}
\caption{a) Sketch of the first four steps of IP dynamics on a Cayley tree
with branching ratio $m=2$. b) Illustration of a causally and geometrically 
connected avalanche whose
size distribution is $P(s)\sim s^{-3/2}$. This characterizes the
stationary state of both IP and the task queue dynamics. }
\label{fig1}
\end{figure}
In our version of the model the most urgent task
is executed with probability $\mu=1$ and replaced in the list by a
constant number $m\ge 2$ of new tasks with random priorities.  All the
features of this model can be clarified by mapping it
into an IP dynamics on a Cayley tree.  A similar mapping on $1-$d IP
has proved to be fruitful also in the case of fixed queue
length \cite{GC07}. Invasion Percolation on a Cayley tree
\cite{IP-Cayley} is defined as follows (see Fig.~\ref{fig1}-a): 
let us take a Cayley tree with
branching ratio $m$ where initially only the top vertex site of the
tree is occupied.  A random number ({\em
fitness}) $x$, extracted from a given probability density function
(PDF) $p(x)$, is assigned once and forever to each empty site (independently
of the others).  At each time-step the site of the growth interface
$\partial {\cal C}_t$ with the highest fitness is occupied. $\partial
{\cal C}_t$ is defined at each time $t$ as the set of empty sites
connected by a first nearest neighbor rule to the connected growing cluster
${\cal C}_t$ of occupied sites up to that time. 
Since for each occupied site other new $m$ sites
enter the growth interface, the number of sites respectively in ${\cal
C}_t$ and in $\partial {\cal C}_t$ at time $t$ are respectively
$\|{\cal C}_t\|=(t+1)$ and $\|\partial {\cal C}_t\|=m+(m-1)t$. Being
the dynamics extremal, the statistical and geometrical features of IP
are independent on the shape of $p(x)$; our choice is to take 
$p(x)=1$ with $x\in [0,1]$.

The exact mapping between IP and our queueing model is done by
identifying sites with tasks, fitness with priority index, growth
interface $\partial {\cal C}_t$ in IP with the task list (i.e. the
queue), and finally the growing IP cluster ${\cal C}_t$ with the set
of executed tasks up to time $t$.

For our purposes we focus on the following features of the asymptotic
stationary state of IP dynamics:\\
1) The distribution (also called {\em normalized interface histogram})
$\phi_s(x)$ of the fitnesses of the interface sites 
(i.e. of the tasks in the queue), has the step-function shape
\[\phi_s(x)=p_c^{-1}\theta(p_c-x)\,,\] 
where $p_c=(m-1)/m$ is the ordinary percolation threshold of the
Cayley tree. This implies that: (i) apart from a vanishing fraction
(i.e. a finite number) of sites, all the interface sites have fitness
$x< p_c$; (ii) since the number of sites in the stationary state is
infinite, only those few sites with $x\ge p_c$ can grow at each
time-step.  Indeed, at each time for the just occupied site (executed
task) $m\ge 2$ new sites (new tasks) enter the interface (queue). This
implies that a fraction $(m-1)/m$ of the sites entered the interface
at any time will never be executed. Since the interface site with
maximal fitness is always executed and the ``fresh'' interface sites
have random fitness, asymptotically only and all the sites with $x\ge
(m-1)/m$ are executed while the others stay forever in the
interface;\\ 2) The cluster of occupied sites is substantially
coinciding with the incipient percolating cluster of ordinary
percolation (i.e. at occupation probability $p=p_c$);\\ 3) The
stationary dynamics self-organizes into a sequence of spatially and
causally connected avalanches of growth activity \cite{Maslov} with a
scale-invariant size distribution $P(s) \sim s^{-3/2}$ 
independently of the value of $m$. Any of these
avalanches (see Fig.~\ref{fig1}-b), say ${\cal A}$, starts with the
growth of a site (the {\em initiator} of ${\cal A}$) with
fitness $x=p_c$ exactly, meaning that all the other
interface sites at that time have $x\le p_c$. Following this
growth, $m$ new sites/tasks ({\em sons}), 
geometrically connected
to the initiator, enter the interface. ${\cal A}$ stops immediately if
all sons have fitness $x< p_c$, and consequently another ``old''
interface site grows with $x=p_c$ [due to the shape of $\phi_s(x)$], 
and therefore initiating a new
avalanche ${\cal B}$, i.e., ${\cal A}$ lasted only one step. If
instead at least one of the sons of the initiator of ${\cal A}$ has
$x>p_c$, then ${\cal A}$ goes on at least one step further as one of
these sons grows. Consequently, other new $m$ ``descendants'' (sons of
a son) of the initiator of ${\cal A}$ enter the interface. Again
${\cal A}$ keeps on if at least one among all the remaining
descendants of any generation (called the {\em avalanche
interface}) has fitness $x>p_c$ otherwise the avalanche stops, and so
on.

The exponent of $P(s)$
can be computed analytically by mapping the avalanche dynamics into a
problem of first return of unbiased random walks. Let $n_t$ be the
number of sites (tasks) with $x\ge p_c=(m-1)/m$ after the $t^{th}$
step of an avalanche (i.e., on its interface).  Since after the growth
of one site $m$ new sites enter the interface, we have the following
Markovian evolution for $n_t>0$
\begin{equation}
n_{t+1} =n_t+j-1\;\; \mbox{with prob.}\; {m \choose j}  p_c^{m-j} (1-p_c)^{j}
\label{n_t}
\end{equation}
with $j=0,1,...,m$.  I.e., $n_t$ follows an ordinary random
walk with independent steps.  As $p_c=(m-1)/m$, the average increment
of $n_t$ in one time-step is zero ({\em martingale} property).
Therefore \cite{redner} the probability distribution of the time $s$
for which $n_s=0$ for the first time (i.e. the duration of the
avalanche) scales as $s^{-3/2}$ at large $s$.
A percolation argument can also be used to find out the same exponent: 
as the avalanche initiator has $x=p_c$ and
the avalanche lasts exactly for a time interval equal to the number of
sites with $x>p_c$ connected to it in the positive time direction,
then the avalanche size is distributed as the finite clusters at the
critical point $p=p_c$ in ordinary percolation on the same tree: 
$P(s)\sim s^{-3/2}$.

Random walk and diffusion theory arguments also permit
to evaluate the stationary state WTD $P_W(\tau)$ for the tasks with $x>p_c$.
We follow here a similar discussion to \cite{GL06}. 
We can write the WTD as
\begin{equation}
P_w(\tau)=\sum_{n=0}^\infty \int_{p_c}^1 dx \tilde{Q}(n,x)G(n,x,\tau)
\label{eq:grin}
\end{equation}
where $\tilde{Q}(n,x)$ is the probability that at a generic time-step
at the stationary state we have exactly $n$ tasks in the queue
(i.e. sites on the IP interface) with priority larger than $x\ge
p_c$.  The quantity $G(n,x,\tau)$ is instead the conditional probability that,
always at the stationary state, a certain task with priority $x \ge
p_c$ added to the list at a time-step when other $n$
tasks with priority larger than $x$ are present, is executed after $\tau$
time-steps.  
We can write the evolution equation for the the number $n_t(x)$ 
of tasks in the list with
priority larger than $x$ at time $t$. Similarly to Eq.~(\ref{n_t}) we
can write for $n_t(x)\ge 1$
\begin{equation}
n_{t+1}(x)=n_t(x)+j-1\;\;\mbox{with prob.}\; {m\choose j}x^{m-j}(1-x)^j\,,
\label{n-t-x}
\end{equation}
where $j=0,1,...,m$.  We can consequently write the master
equation for the probability $Q(n,x,t)$ that at time $t$ there are
exactly $n$ tasks in the list with priority larger than $x$. 
To aim of simplicity let us write it for $m=2$ for which $p_c=1/2$.
From Eq.~(\ref{n-t-x}) we can write for $n\ge 3$
\begin{eqnarray}
Q(n,x,t+1)&=&Q(n+1,x,t)x^2+Q(n,x,t)2x(1-x) \nonumber\\
&+&Q(n-1,x,t)(1-x)^2
\label{hier}
\end{eqnarray}
while for $n\le 2$ we have
\begin{eqnarray}
Q(2,x,t+1)&=&Q(3,x,t)x^2+Q(2,x,t)2x(1-x)\nonumber\\ 
          &+&Q(1,x,t)(1-x)^2+Q(0,x,t)(1-x)^2\nonumber\\ 
Q(1,x,t+1)&=&Q(2,x,t)x^2+Q(1,x,t)2x(1-x)\nonumber\\ 
          &+&Q(0,x,t)2x(1-x)\;;\;\nonumber\\
Q(0,x,t+1)&=&Q(1,x,t)x^2+Q(0,x,t)x^2 
\label{hier2}
\end{eqnarray}
$\tilde Q(n,x)$ is given by the stationary solution of the above equations.  
In order to find both $\tilde Q(n,x)$ and
$G(n,x,\tau)$ we can now proceed in a similar way to \cite{GL06}. It
is simple to show that the well-normalized stationary solution for 
$x\ge p_c=1/2$ of Eqs.~(\ref{hier}) and (\ref{hier2}) is
\begin{eqnarray}
\label{stat-q}
&&\tilde Q(n,x)=\frac{2(x-p_c)}{x^2}\left[\frac{(1-x)^2}{x^2}\right]^{n-1}
\;\;\mbox{for}\; n\ge 2\\
&&\tilde Q(1,x)=2\frac{1-x^2}{x^2}(x-p_c)\;;\;
\tilde Q(0,x)=2(x-p_c)\,.\nonumber
\end{eqnarray}
Note that for $x\to p_c^-$ any $\tilde Q(n,x)\to 0$
with the ratio $\tilde Q(n,x)/\tilde Q(l,x)\to 1$ for any $n,l\ge 2$,
i.e., the distribution of the number $n_{t\to\infty}(p_c)$ 
becomes practically uniform.

The quantity $G(n,x,\tau)$ can be found by Eq.~(\ref{hier})
in complete analogy with \cite{GL06} and \cite{GL08} leading
both to the same correct scaling behavior $P_W(\tau)\sim \tau^{-3/2}$.
We refer here to the former as it is of simpler formulation.
First of all we note that Eq.~(\ref{hier}), in both the continuous time 
and $n=y$ approximation, becomes the diffusion equation:
\begin{equation}
\partial_t Q(y,x,t)=c(x)\partial_y^2Q(y,x,t) +d(x)\partial_y Q(y,x,t)\,,
\label{q-cont}
\end{equation}
with $c(x)=x^2$ and $d(x)=x^2-(1-x)^2$. Since we are
considering $x\ge p_c=1/2$ we have $d(x)\ge 0$, i.e., there is a drift
to the small $y$ (i.e. $n$) direction.  $G(n,x,\tau)$ can be
seen as the probability that at the stationary state, fixed $x$ and given
that at time $t=0$ it is $y=n$, one has $y=0$ for the first at time $t=\tau$.
This implies that \cite{redner,GL06}
\[G(n,x,t)=-\partial_t\left[\int_0^\infty dy\,Q(y,x,t)\right]\,,\]
where here $Q(y,x,t)$ is the solution of 
Eq.~(\ref{q-cont}) with initial condition $Q(y,x,0)=\delta(y-n)$.
All this gives
\[G(n,x,\tau)=\frac{n}{\sqrt{4\pi c(x)}t}\exp\left\{-\frac{[n-d(x)t]^2}{4c(x)t}
\right\} \] 
We now use this result and
Eq.~(\ref{stat-q}) in Eq.~(\ref{eq:grin}) to find
$P_W(\tau)$.  It is simple \cite{GL06} to show
that for large $\tau$ we have $P_W(\tau)\sim \tau^{-3/2}$. In other
words each task with $x\ge p_c$ has to wait a finite portion of the
avalanche duration before being executed.  Note that all these results are
completely independent of the integer branching factor
$m>1$. From Eq.~(\ref{n-t-x}) it is natural to expect to have the
same result in the case in which at each time step $m$ is not constant
but fluctuates with independent fluctuations such that
$\left<m\right>\ge 1$ and finite variance.  This is
the reason why our model share the same statistical features with that in 
\cite{GL06}.  In the case where
$\left<m^2\right>=+\infty$ we expect anomalous exponents for both
$P(s)$ and $P_W(\tau)$ as the random walks Eqs.~(\ref{n_t}) and
(\ref{n-t-x}) become Levy flights as shown in \cite{masuda}.


We now address the question on how fast is the approach to
stationarity in such models.  Again some rigorous theoretical results
for IP on a tree turn to be very useful to this end. We summarize here
the main results in literature, and then propose a simple mean-field approach
showing how slow the relaxation to the right stationary state is.  
In \cite{IP-Cayley} the main exact result, adapted to our
notation, states that the probability that at time $t$ of the dynamics
a task with priority smaller than $(p_c-\epsilon)$ is executed,
vanishes exponentially fast for large $t$ for $\epsilon>0$, but as
$t^{-1/2}$ for $\epsilon\to 0^+$. This suggests that deviations from
the stationary dynamics disappear as $t^{-1/2}$.  In \cite{IP-tree}
instead it is shown rigorously that: (i) IP on a Cayley tree has in
the infinite time limit a unique backbone. In terms of the task
dynamics this means that there is a unique infinite chain of executed
task which are causally connected in the IP sense above.  (ii) The
minimal priority of the executed tasks staying on the backbone beyond
the $k^{th}$ generation of the Cayley tree (see Fig.~\ref{fig1}-a)
is $p_c(1-Z/k)$ for large $k$ where $Z$ is an exponential 
random variable with unitary mean.

We now present a simple mean field argument showing this slow approach
of the list dynamics to the right stationary state.  We study the
dynamics of the above introduced normalized distribution $\phi(x,t)$
of the priorities of the tasks in the queue (fitness {\em histogram} of
interface sites in IP) at time $t$. 
In order to write a closed equation for $\phi(x,t)$ we use the Run
Time Statistics (RTS) which is a probabilistic method introduced to
describe IP and related dynamics, and evaluate the statistical weight
of all different ``histories'' of the dynamics (i.e. paths in the
realization space) \cite{RTS1,RTS2,RTS3}.  Let $h(x,t)dx$ be the
number of tasks in the queue at time $t$ with
priority in $[x,x+dx]$ in a single realization. We can write rigorously:
\begin{equation}
h(x,t+1)=h(x,t)-m(x,t+1)+m\,,
\label{histo1}
\end{equation} 
where $m(x,t)$ is the PDF of the priority of the executed task at time
$t$ conditional to the whole past history. By calling $p_i(x,t)$ the
PDF of the priority of the $i^{th}$ task in the queue at time $t$
conditional to the same past history, and assuming that the executed
task at that time is the $j^{th}$, a good approximation for $m(x,t+1)$
\cite{RTS2} is $m(x,t+1)=\frac{1}{\mu_j(t)}p_j(x,t)\prod_{i(\ne
j)}^{\partial {\cal C}_t}\left[ \int_0^x dy\,p_i(y,t)\right]$, where
$\mu_j(t)=\int_0^1 dx\,p_j(x,t) \prod_{i(\ne j)}^{\partial {\cal
C}_t}\left[\int_0^x dy\,p_i(y,t)\right]$ is the probability of
selecting $j$ conditional to the past history.  We now average
Eq.~(\ref{histo1}) over all the possible realizations of the dynamics
up to time $(t+1)$ using the symbol $\left<\cdot\right>_{t+1}$ for
this average.  By definition we have
$\left<h(x,t+1)\right>_{t+1}=(m-1)(t+1)\phi(x,t+1)$ and
$\left<h(x,t)\right>_{t+1}=\left<h(x,t)\right>_t=(m-1)t\phi(x,t)$.
In order to take the same average of $m(x,t+1)$
note that, if $A(i_0,i_1,...,i_{t-1}, i_t)$ is a function of the queue
history up to time $(t+1)$ identified by the sequence of executed
tasks $\{i_0,i_1,...,i_{t-1},i_t\}$, by the rules of conditional
probability, we can write
\[\left<A(i_0,i_1,..., i_t)\right>_{t+1}=
\left<\sum_{j\in \partial{\cal C}_t}\mu_j(t)A(i_0,i_1,...,i_{t-1},
j)\right>_t \,.\] 
We therefore have
\[\left<m(x,t+1)\right>_{t+1}=\left<\sum_{j\in 
\partial{\cal C}_t}p_j(x,t)\prod_{i(\ne j)}^{\partial {\cal C}_t}
\left[\int_0^x dy\,p_i(y,t)\right]\right>_t\,.\] 
Considering that by definition $\left<p_j(x,t)\right>_t=\phi(x,t)$, we
now introduce the mean field approximation consisting in replacing the
average of the above products of $p_l(x,t)$ with the products of the
averages, i.e., \
\be
\left<m(x,t+1)\right>_{t+1}=(m-1)t\phi(x,t)\left[\int_0^x dy\,
\phi(y,t)\right]^{(m-1)t-1}\,.
\label{mf}
\ee 
We can now write the mean-field equation for $\phi(x,t)$ as
\bea \phi(x,t+1)&=&\frac{t}{t+1}\phi(x,t)\left\{1-\left[\int_0^x dy\,
\phi(y,t)\right]^{(m-1)t-1}\right\}\nonumber\\ &+&\frac{m}{(m-1)(t+1)}
\label{phi}
\eea 
This strong decorrelating approximation is expected to
lead to a faster relaxation to stationarity than the actual one. 
We show however that, even in this approximation, the 
stationary state is the right one and the approach to it is power
law. Integrating both sides of Eq.~(\ref{phi}), taking the continuous
time limit and $t\gg 1$ we get 
\be 
\partial_t\psi(x,t)={-1 \over
t+1}\left[\psi+\frac{1}{m-1}\psi^{(m-1)t}-\frac{mx}{m-1}\right]
\label{phi2}
\ee
where $\psi(x,t)=\int_0^x dx'\,\phi(x',t)$ is the cumulative average priority 
distribution, and we have assumed $\psi(0,t)=0$ at all $t$. 
The initial condition for Eq.~(\ref{phi2}) is $\psi(x,0)=x$. 
Since $\phi(x,t)$ is a normalized PDF, we have $\psi(x,t)\ge 0$, 
non-decreasing in $x$ and $\psi(1,t)=1$.

In the $x$ region were $(1-\psi)\gg 1/[(m-1)t]$ we can approximate 
Eq.~(\ref{phi2}) simply with 
\be
\partial_t\psi(x,t)=-{1 \over t+1}\left(\psi-\frac{mx}{m-1}\right)\,,
\ee
which leads to the solution for sufficiently large $t$ 
\be \psi(x,t)\!\!=\!\!
\frac{mx}{m-1}\!\left(\!1-\frac{1}{mt}\right)\;\mbox{for }x<{m-1\over m}
-\frac{1}{m^2t}\,.
\label{phi3}
\ee
Note that $p_c=(m-1)/m$.  Moreover in the $x$ region were
$\epsilon=(1-\psi)\ll 1/[(m-1)t]$ it is simple to show that the
following approximation holds
\be
\partial_t \epsilon(x,t)= -\epsilon(x,t)+\frac{m(1-x)}{(m-1)(t+1)}
\ee
whose solution is $\psi(x,t)=1-\epsilon(x,t)$ with 
\be
\epsilon(x,t)\!=\!\frac{m(1-x)}{(m-1)(t+1)}
\left[1\!+O\left(\!\frac{1}{t}\!\right)\right]\;\label{phi4}
\ee
when $ x\gg (m-1)/m$. 
All this means that
\[
\phi(x,t)=\frac{\theta(p_c-x)}{p_c}+\delta\phi(x,t)
\]
with $\delta\phi(x,t)\sim 1/t$. Therefore even in this mean
field approximation, for which we expect a faster relaxation, 
deviations from it vanish as slowly as $1/t$.

In conclusion, we have shown a way to analytically compute all the
main features of the Barab\'asi model of human dynamics with
time-increasing queue length. This is done by
using Invasion Percolation on a Cayley tree and random walk theory.  We
believe that the approach we introduced, allows us to
describe quantitatively two intuitive features of tasks queues.  The
first feature is that some tasks seem to remain indefinitely before
being processed; secondly we recover naturally the fact that in real
world execution of a task has often the effect to generate an
avalanche of new tasks.  Through our approach one can study both the
stationary state dynamics and the approach to it. This shows that both
are characterized by temporal power laws as typical for extremal
dynamics in quenched disorder \cite{RTS1,RTS2,RTS3}.


\end{document}